%% file: ifacconf.tex
\documentclass{ifacconf}
\usepackage{amsmath,amssymb,amsfonts}
\usepackage{algorithm}
\usepackage{algorithmic}
\usepackage{enumitem}
\usepackage{graphicx}
\usepackage{cuted}
\usepackage{color}
\usepackage{flushend}
\usepackage{mathtools}
\usepackage{array}
\usepackage{subcaption}
\mathtoolsset{showonlyrefs}
\usepackage{multirow}
\usepackage{braket}
\def\Gc{{\mathcal G}}
\def\Vc{{\mathcal V}}
\def\Ec{{\mathcal E}}
\def\Nc{{\mathcal N}}

\usepackage{xcolor}
\newcommand{\SCB}[1]{\textcolor{black}{#1}}
\usepackage{graphicx}      
\usepackage{natbib}        
\begin{document}
\begin{frontmatter}
\title{Security of Gradient Tracking Algorithms Against Malicious Agents\thanksref{footnoteinfo}} 
%
\thanks[footnoteinfo]{This work is supported by the Swedish Research Council under the grant 2024-00185.}
\author[First]{Sribalaji C. Anand} 
\author[Second]{Alexander J Gallo}
\author[First]{Nicola Bastianello} 
\address[First]{School of Electrical Engineering and Computer Science, KTH Royal Institute of Technology, Sweden. (email: \{srca, nicolba\}@kth.se)}
\address[Second]{Department of Electronics, Information and Bioengineering, Politecnico di Milano, Milano, Italy. \\(e-mail: alexanderjulian.gallo@polimi.it)}
\begin{abstract}                
Consensus algorithms are fundamental to multi-agent distributed optimization, and their security under adversarial conditions is an active area of research. While prior works primarily establish conditions for successful global consensus under attack, little is known about system behavior when these conditions are violated. This paper addresses this gap by investigating the robustness of the Wang--Elia algorithm, which is a robust to noise version of gradient tracking algorithm, in the presence of malicious agents. We consider a network of agents collaboratively minimizing a global cost function, where a subset of agents may transmit faulty information to disrupt consensus. To quantify resilience, we formulate a security metric as an optimization problem, which is rooted in centralized attack detection literature. We provide a tractable reformulation of the optimization problem, and derive conditions under which the metric becomes unbounded, identifying undetectable attack signals that reveal inherent vulnerabilities. To facilitate design and analysis, we propose a well-posed variant of the metric and propose design methods to enhance network robustness against stealthy adversarial attacks. Numerical examples demonstrate the effectiveness of the proposed framework to enhance the resilience of multi-agent distributed optimization.
\end{abstract}
\end{frontmatter}
%
\section{Introduction}\label{Intro}
\input{Introduction.tex}
%
%
\section{Problem Description}\label{Sec:PF}
\input{2_Problem_formulation.tex}
\section{Undetectable attacks}\label{sec:undetec}
\input{3_UDA.tex}
\section{Quantifying security against undetectable attacks}\label{sec:MOOG}
\input{4_modify_OOG.tex}
\section{Design problem}\label{sec:design}
\input{5_Design.tex}
\section{Extension to nonlinear systems}\label{sec:discuss}
\input{6_Discussion.tex}
\section{Conclusion}\label{sec:con}
In this paper, we introduced a security metric to assess the robustness of the Wang--Elia algorithm against adversarial attacks. We first identified a class of undetectable attacks, revealing fundamental vulnerabilities in the system. To address the ill-posedness of the original metric, we proposed a well-posed variant by introducing artificial delays into the performance output. Numerical studies demonstrated the effectiveness of the framework in enhancing robustness, and we applied the metric for design purposes to improve the resilience of network graphs against malicious agents. Future directions include extending the framework beyond polynomial nonlinear objectives.
\bibliography{ifacconf}                
\end{document}

%% file: Introduction.tex
In many modern applications, such as smart grids, sensor networks, and federated learning, data and computational resources are inherently distributed across multiple agents \citep{rauniyar2023federated}. A multi-agent setup also avoids the need for a central controller to collect all data, reducing communication costs, improving scalability, and preserving data privacy. Most of the tasks that multi-agent systems need to solve can be encoded as consensus and/or distributed optimization (DO) problems. 

The security of DO algorithms has been extensively studied by \citet{ishii2022overview,pirani2023graph,liu2024survey} and references therein. Existing works typically establish necessary conditions under which an algorithm can successfully recover the global optimum, provided these necessary conditions are satisfied. However, less attention has been devoted to the complementary question: \emph{what if the necessary conditions are not met?} In such cases, can we quantify the robustness of the distributed network? This paper focuses precisely on this issue. 

In particular, we investigate the security of the Wang--Elia algorithm \citep{wang2010control}, which is a robust-to-noise gradient tracking algorithm. The algorithm has gained wide adoption in the literature due to its favorable convergence properties against additive perturbations \citep{bin2022stability}. In particular, when the perturbations are bounded in magnitude, the algorithm is guaranteed to converge linearly to the global optimum.

In this paper, we study the security of Wang--Elia algorithm against malicious agents. In particular we consider $N$ agents optimizing a global cost function $f(x) = \sum_{i=1}^{N} f_i(x)$, and each agent $i$ has access to its local objective function $f_i(x)$, and the value of $x$ from its neighbors. We consider that one of the agents sends malicious information to its neighbors with the objective of disrupting the consensus. In order to counteract the attack, one of the agents is monitored. Given the above problem setup, we present the following contributions.
\begin{enumerate}[label=(\roman*)]
    	\item We formulate an optimization problem whose optimal value serves as a security metric, quantifying the robustness of the DO algorithm against malicious agents. Since this problem is non-convex, we propose an equivalent semidefinite program (SDP) that enables tractable analysis. 
   	\item If the value of the security metric is unbounded, it implies the existence of undetectable attack signals, thereby revealing a fundamental system vulnerability. To this end, we derive explicit conditions under which the security metric becomes unbounded, and we characterize the corresponding attack signals. \item When the security metric is unbounded, it cannot be used for design. To resolve this, we propose a well-posed, amended version of the metric.
    	\item Finally, we propose two design methods to enhance the resilience against malicious stealthy agents. 
\end{enumerate}

Security metrics have been extensively studied for consensus algorithms \citep{anand2025quantifying}. For instance, \cite{nguyen2024security} employs a security metric but lacks a method to address its ill-posedness. \cite{arnstrom2025efficiently} consider energy-bounded attacks, which may not be realistic due to the potential unavailability of the bound. Other works, such as \cite{yuan2019stackelberg,shukla2022robust}, adopt a stackelberg game approach but do not include stealthy attacks \citep{umsonst2023bayesian}. {The metric proposed in this paper is related to the output-to-output gain (OOG) introduced in \cite{teixeira2015strategic}. While the ill-posedness of OOG was partially addressed in \cite{anand2020joint}, we provide a comprehensive analysis and extend the metric to a distributed setting.}

\emph{Outline:} The remainder of the paper is organized as follows. {In Section~\ref{Sec:PF}, we introduce a security metric for DO problems and formalize the problem studied}. In Section~\ref{sec:undetec}, we employ the security metric to characterize undetectable attacks. Section~\ref{sec:MOOG} presents an amended version of the security metric that ensures well-posedness, and Section~\ref{sec:design} leverages this amended metric to address the design problem. The effectiveness of the proposed methods is demonstrated through numerical examples provided throughout the paper. Finally, Section~\ref{sec:con} offers concluding remarks. 

\emph{Notation:} 
%
%
%
%
%
%
{Given a differentiable function $f(x): \mathbb{R}^n \to \mathbb{R}$, $\nabla_x f(x)$ represents the derivative of the function $f(x)$ with respect to $x$. Let $\Gc \triangleq (\Vc, \Ec, K)$ be a weighted undirected graph with the set of $N$ vertices $\Vc = \{v_1, v_2,...,v_N\}$, the set of edges $\Ec \subseteq \Vc \times \Vc $, and the weighted matrix $K$. An edge of $\Gc$ is denoted by a pair $(v_i,v_j) \in \Ec$.} 

%% file: 2_Problem_formulation.tex
This section presents the DO framework, adversarial model, and problem formulation.
\subsection{Distributed optimization under malicious agents}
Consider an optimization problem of the form 
\begin{equation}\label{eq:main:dist:opt}
\min_{x \in \mathbb{R}^n} F(x) \triangleq \sum_{i=1}^{N} f_i(x),\; f_i(x) \triangleq \frac{1}{2}x^TQ_ix + c_i^Tx.
\end{equation} 
where $Q_i \succeq 0$, and ${x^* \in \mathbb{R}^n}$ denotes the optimal argument of \eqref{eq:main:dist:opt}. We consider that the optimization problem \eqref{eq:main:dist:opt} is solved in a distributed fashion as described next. 

Consider an undirected, connected graph $\Gc = (\Vc,\Ec,K)$ with $N$ agents. Each agent has an objective function $f_i(x)$, only known to agent $i$. The objective of the agents is to find the value of $x^*$, that solves \eqref{eq:main:dist:opt}. To this end, each agent $i \in \{1,\dots,N\}$ updates its estimate of $x$ using the Wang--Elia algorithm \cite[(6)]{bin2022stability} given as follows 
\begin{equation}\label{eq:Wang:single}
\begin{aligned}
        x_i[k+1] &= x_i[k] + \sum_{j\in \Nc_i} k_{ij} ( {x}_j[k] - x_i[k] + {z}_j[k]  \\
        & \hspace{30pt}- z_i[k] ) - \alpha \nabla_x f_i(x_i[k]) + b_{i} {a[k]} \\
    z_i[k+1] &= z_i[k] - \sum_{j\in \Nc_i} k_{ij} \left( {x}_j[k] - x_i[k] \right) + b_{i} {a[k]}
\end{aligned}
\end{equation}
where ${z \in \mathbb{R}^n}$ is an auxiliary state, $\begin{bmatrix}
    x_i[0]\\
    z_i[0]
\end{bmatrix} = \begin{bmatrix}
    x_{i,0}\\
    z_{i,0}
\end{bmatrix} \forall i$, $\alpha$ is the step size chosen according to Theorem 1 in \cite{bin2022stability}, $a[k] \in \mathbb{R}^{n}$ is the attack vector injected by agent $i$. If agent $i$ is malicious, then $b_{i} = 1$ and $b_{i} = 0$ otherwise. 

We consider that one of the $N$ agents is malicious and aims to disrupt the consensus. The malicious agent is also called an adversary in the reminder of this article. Then, using \eqref{eq:main:dist:opt}, the aggregated version of \eqref{eq:Wang:single} can be written as 
\begin{equation}\label{eq:Wang:agg}
\bar{x}[k+1] \triangleq 
\begin{bmatrix}
    x[k+1]\\
    z[k+1]
\end{bmatrix} = \bar{A} \bar{x}[k] + \bar{B}a[k] + \bar{c}  
\end{equation}
where $ \bar{A} \triangleq \begin{bmatrix}
I-K-\alpha Q  & -K\\
K & I 
\end{bmatrix}, \bar{c} \triangleq \begin{bmatrix}
    c\\0
\end{bmatrix}, {\bar{B} \triangleq \begin{bmatrix}
    	B \\
  	B
\end{bmatrix}}$, and $x[k]$, $z[k]$, $c$, and $B$ are the stacked versions of $x_i[k]$, $z_i[k]$, $c_i[k]$, and \SCB{$b_{i}I_n$} respectively, and $Q = \text{diag}\left[ Q_1 \dots Q_N \right]$. The matrix $K$ is the graph adjacency matrix that satisfies $K=K^T, \; \mathbf{1}_{N}^T K = \mathbf{0}_{N}^T, \; \rho(K) \in [0,\; 1).$

To detect attacks, the operator monitors the evolution of the states in one of the agents. Given that agent \( m \in \{1, \dots, N\} \) is monitored, its output is defined as
\begin{equation}
    \bar{y}_m[k] = \bar{C}_m \bar{x}[k] - \kappa,
\end{equation}
where {\(\bar{C}_m = W\check{C}_m \in \mathbb{R}^{2n \times 2nN}\). Here the matrix $\check{C}_m$ extracts the states \(x_m[k]\) and \(z_m[k]\) from \(\bar{x}[k]\), and $W = \operatorname{diag}\!\big((1 - w_i) I_n,\, w_i I_n\big)$ is a weight matrix chosen by the operator, and $w_i \in [0,\;1],\;\forall i \in \{1,\dots,n\}$ is a tuning parameter}. For instance, if $w_i=1$ ($w_i=0$), then the monitor ignores the evolution of $x_m$ ($z_m$). The constant \(\kappa\) is designed apriori such that, under nominal conditions, $\lim_{k \to \infty} \bar{y}_m[k] \approx 0,\; \forall \bar{x}[0]$. The monitoring agent raises an alarm if the following condition holds
\begin{equation}
\left\Vert \bar{y}_m[k]\right\Vert_{\ell_2, [1,L]}^2 > \epsilon \implies \text{alarm}
\end{equation}
where $\epsilon \in \mathbb{R}^{+}$ is the tunable detection threshold. We describe an approach to choose the detector threshold $\epsilon$ and $\kappa$ in Example~\ref{exmp:2} and \ref{exmp:1}, respectively.
\begin{rem}
\SCB{We consider the case of a single agent under attack, which simplifies the analysis without loss of generality. The results presented readily extend to multiple attack nodes. As coordinated attacks often enhance stealthiness, our analysis focuses on this stealthiness aspect.}$\hfill \triangleleft$
\end{rem}
\begin{rem}\label{rem:attack:different}
\SCB{We consider that the same attack vector $a[k]$ is injected into both states $x$ and $z$. The results presented extend to distinct attack vectors for $x$ and $z$, which would modify the attack input matrix $\bar{B}$; the main conclusions still hold, though the attack becomes more powerful.}$\hfill \triangleleft$
\end{rem}
\subsection{Adversarial model}
Next, we discuss the adversary under consideration.
\subsubsection{Adversarial objective:} The goal of DO problem is to have ${y}_{p,i} \triangleq \left( x_i-x_{i+1}\right) \to 0 \;\forall \; i$. Let $\bar{y}_p$ represent the stacked vector of $y_{p,i}$ and is given by 
\begin{equation}
    \bar{y}_p[k] = \bar{C}_p \bar{x}[k],
\end{equation}
where $\bar{C}_p$ is a matrix which computes the error signal from $\bar{x}[k]$, and $\bar{x}[k]$ is obtained from \eqref{eq:Wang:agg}. Then we consider an adversary that aims to increase the value of $\Vert \bar{y}_{p} \Vert_{\ell_2,[1,L]}^2$ which, when non-zero, denotes a loss of optimality. 
\subsubsection{Adversarial constraints:} As mentioned before, the operator monitors the evolution of the states in one of the agents to detect attacks. The adversary injects a stealthy  attack so that it does not raise an alarm. To provide the formal definition of stealthy attacks, let the closed-loop system from the attack input to the performance and detection output be given by 
\begin{equation}\label{eq:CL}
    \begin{aligned}
        \bar{x}[k+1] &= \bar{A} \bar{x}[k] + \bar{B}a[k] + \bar{c},\\
        \bar{y}_p[k] &= \bar{C}_p \bar{x}[k],\quad \bar{y}_m[k] = \bar{C}_m \bar{x}[k] - \kappa
    \end{aligned}
\end{equation}
Then a stealthy attack can be defined as follows.
\begin{definition}[Stealthy attack]
    An attack vector $a[k]$ injected into \eqref{eq:CL} is defined to be stealthy if 
    \begin{equation}\label{eq:stealthy}
\left\Vert \bar{y}_m[k]\right\Vert_{\ell_2, [1,L]}^2
\leq \epsilon,
\end{equation}
where $\epsilon \in \mathbb{Z}^{+}$ is a predefined constant. $\hfill \triangleleft$
\end{definition}
\subsubsection{Adversarial knowledge:} The adversary knows the topology $\Gc$, the matrices $Q_i, \alpha_i,$ and $c_i, \forall i \in \{1,\dots,N\}$. In reality, it might be hard for the adversary to obtain this knowledge; however, such setups helps us analyze the worst-case. We next formulate the problem studied. 
\subsection{Problem Formulation}
As discussed previously, the objective of the adversary is to inject a stealthy attack vector $a$ that maximizes the state discrepancy among the agents. Formally, this corresponds to the adversary solving the following non-convex optimization problem
\begin{equation}\label{eq:OOG}
\gamma \triangleq \sup_{a \in \ell_{2e}} \left\{\; \Vert \bar{y}_p \Vert_{\ell_2, [1, L]}^2 \vert \; \eqref{eq:CL}, \eqref{eq:stealthy} \right\}
\end{equation}
where $\gamma$ quantifies the maximal degradation in consensus that can be induced by a stealthy attack. In the following, we refer to the optimization problem \eqref{eq:OOG} as the \emph{security metric}. The value of $\gamma$ provides a measure of the robustness of the Wang--Elia consensus protocol: a small $\gamma$ indicates that the network is robust to stealthy attacks, whereas a large $\gamma$ suggests vulnerability. Next, we derive the conditions under which the security metric \eqref{eq:OOG} is bounded.  

%% file: 3_UDA.tex
In this section, we aim to discuss the existence of stealthy attacks that do not raise alarms, and can cause unbounded deviation in the consensus. To this end, consider the dynamical system \eqref{eq:CL} whose output can be written as follows, thanks to linearity
\begin{equation}
   \begin{aligned}
   \bar{y}_{p} &= \bar{y}_{p,n} + \bar{y}_{p,a},\;
    \bar{y}_{m} = \bar{y}_{m,n} + \bar{y}_{m,a}\\
    \text{where}\;\;
        \bar{x}_n[k+1] &= \bar{A} \bar{x}_n[k] + \bar{c}, \bar{x}_n[0]=\bar{x}_0\\
        \bar{y}_{p,n}[k] &= \bar{C}_p \bar{x}_n[k],\quad        \bar{y}_{m,n}[k] = \bar{C}_m \bar{x}_n[k] - \kappa\\
        \bar{x}_a[k+1] &= \bar{A} \bar{x}_a[k] + \bar{B}a[k], \bar{x}_a[0]=\mathbf{0}\\
        \bar{y}_{p,a}[k] &= \bar{C}_p \bar{x}_a[k],\quad        \bar{y}_{m,a}[k] = \bar{C}_m \bar{x}_a[k] 
    \end{aligned}
\end{equation}
where the subscript $n$ denotes the nominal output and the subscript $a$ denotes the attacked outputs. It follows that, if there exists a non-zero attack vector that yields $\bar{y}_{m,a} =0 $, but yields $\bar{y}_{p,a} \neq 0$, then it is an undetectable attack. {For such attacks, the value of \eqref{eq:OOG} becomes unbounded since there exists an attack $a \in \ell_{2e}$ driving $\|\bar{y}_{p,a}\|_{\ell_2,[1,L]}^2 \to \infty$ while \eqref{eq:stealthy} is trivially satisfied. In this section, we aim to formalize the conditions under which such attacks exist.} Before we present the results, we introduce the following definition from \cite{teixeira2015secure}. 
\begin{definition}[Unstable Invariant Zero]
Given a strictly proper system $\Sigma = (A,B,C,0)$ where $A \in \mathbb{R}^{n \times n}$, $B \in \mathbb{R}^{n \times m}$, and $C \in \mathbb{R}^{p \times n}$, $\lambda \in \mathbb{C}$ is an invariant zero of $\Sigma$ if $\exists\;x_0 \in \mathbb{R}^n, g \in \mathbb{R}^m$
\begin{equation}\label{eq:zero}
    \begin{bmatrix}
        \lambda I-A & B\\
        C & 0
    \end{bmatrix}\begin{bmatrix}
        x_0\\ g
    \end{bmatrix} = \begin{bmatrix}
        0 \\ 0
    \end{bmatrix}.
\end{equation}
The invariant zero is defined to be unstable if $\vert \lambda \vert >1$.$\hfill\triangleleft$
\end{definition}
\begin{definition}[Relative degree]
Given a strictly proper system $\Sigma = (A,B,C,0)$, the relative degree of $\Sigma$ is denoted by $\delta$, and is given by $\delta \triangleq 1+\min \{ j \;| CA^jB \neq 0, j \geq 0 \}$. $\hfill\triangleleft$
\end{definition}

In words, $\delta$ denotes the amount of time delay before the input signal affects the output. Using this definition of relative degree, we present the main result. 
\begin{thm}\label{thm:1}
The value of the security metric \eqref{eq:OOG} is unbounded if one of the following conditions holds
\begin{enumerate}[label=(\roman*)]
    \item The tuple $\Sigma_m = (\bar{A}, \bar{B}, \bar{C}_m,0)$ has an unstable invariant zero which is \emph{not} an unstable invariant zero for the tuple $\Sigma_p = (\bar{A}, \bar{B}, \bar{C}_p,0)$
    \item The relative degree of $\Sigma_m$ is greater than that of $\Sigma_p$.
\end{enumerate}
\end{thm}
\begin{pf}
For any fixed horizon $L \in \mathbb{Z}^{+}$, it follows that $\bar{\mathbf{y}}_{p,a} = \mathbf{O}_{ap} \mathbf{a},\bar{\mathbf{y}}_{m,a} = \mathbf{O}_{am} \mathbf{a},$
where the vectors in bold denote the time-stacked vectors, and the matrices are Toeplitz. We aim to show that, when either of the theorem conditions hold, the value of \eqref{eq:OOG} is unbounded. Equivalently, we can show that when the theorem conditions hold, $\exists \; \mathbf{a} \neq 0$ s.t. $\mathbf{y}_{m,a} = 0$, $\mathbf{y}_{p,a} \neq 0$. 

{Proof of sufficiency of $(i)$}: Let $\lambda_{m}$ be an invariant zero of $\Sigma_m$, and let $\exists x_{0,m}$ and $g_m$ such that \eqref{eq:zero} holds. Let $\bar{x}[0] = \bar{x}_{0,m}$, and let us consider an attack input of the form $a[k] = \lambda_m^k\;g_m$. Then by definition of zero dynamics attack, it holds that $\mathbf{y}_{m,a} =0$ \citep{teixeira2015secure}. Since $z$ is not an unstable zero of $\Sigma_p$ (from the theorem statement $(i)$), it also holds that $\mathbf{y}_{p,a} \neq 0$. 
This concludes the proof of $(i)$. {We assume $\bar{x}[0] = x_{0,m}$, so the adversary must first drive the system to this initial state before injecting the attack $a[k] = \lambda_m^k g_m$. During this phase, the detector output energy may increase; however, it will not trigger an alarm if $x_{0,m}$, and $g_m$ are scaled down appropriately.}

{Proof of sufficiency of $(ii)$}: Let us denote the relative degree of $\Sigma_p$ and $\Sigma_m$ as $\delta_p$ and $\delta_m$ respectively. Let $\delta_m > \delta_p$. Then for any horizon $L$, let the adversary construct an attack of the form \eqref{eq:attack:rd} where $\beta$ is a term of arbitrary magnitude. The non-zero attack $\beta$ is applied at the time instant $L-\delta_m$. Since the relative degree of $\Sigma_m$ is $\delta_m$, the attack vector does not affect the measurement output. However since $\delta_m > \delta_p$, the non-zero attack term influences $\mathbf{y}_{p,a}$. Thus, the attack \eqref{eq:attack:rd} drives $\|\bar{y}_{p,a}\|_{\ell_2,[1,L]}^2 \to \infty$ while \eqref{eq:stealthy} is trivially satisfied. 
This concludes the proof. $\hfill \blacksquare$
\end{pf} 
The result in Theorem~\ref{thm:1} implies that when the conditions of Theorem~\ref{thm:1} hold, an adversary can cause unbounded deviation in the consensus protocol; exposing a limitation of the algorithm proposed in \citet{wang2010control}. Additionally, when condition $(i)$ in Theorem~\ref{thm:1} holds, the adversary injects an attack of the form 
\begin{equation}\label{eq:attack}
a[k] = \lambda_m^k g_m,    
\end{equation}
where $\lambda_m$ is the invariant zero of $\Sigma_m$. Similarly, when condition $(ii)$ in Theorem~\ref{thm:1} holds, the adversary injects
\begin{equation}\label{eq:attack:rd}
    a[k] = \begin{cases}
        0_{n \times 1} & 0 \leq k \;{\leq}\; L-\delta_m\\
        \beta \otimes 1_{n \times 1} & L-\delta_m {<} \;k \leq L
        \end{cases}
\end{equation}
where $L \in \mathbb{Z}^{+}$ is the horizon length, and $\delta_m$ is the relative degree of $\Sigma_m$, and $\beta$ is an arbitrary constant. Thus, to cause unbounded performance loss, the adversary injects an attack signal of growing magnitude; although the attack vector can be scaled. We depict the performance loss caused by such attacks in Example~\ref{exmp:1}.

Finally, Theorem~\ref{thm:1} exposes a limitation of the security metric \eqref{eq:OOG}. For instance, if we want to design the step size $\alpha$ or the detector threshold $\epsilon$ to minimize the security metric, the value of the metric should always be bounded. In other words, the security metric \eqref{eq:OOG} should be well-posed. In the next section, we propose a method to amend the security metric \eqref{eq:OOG} thus making it well-posed, and thus a useful tool to tune $\alpha$ or $\epsilon$ for increased robustness.

\begin{figure*}[t]
    \centering
    \begin{subfigure}{0.32\textwidth}
        \centering
        \includegraphics[width=1.1\linewidth]{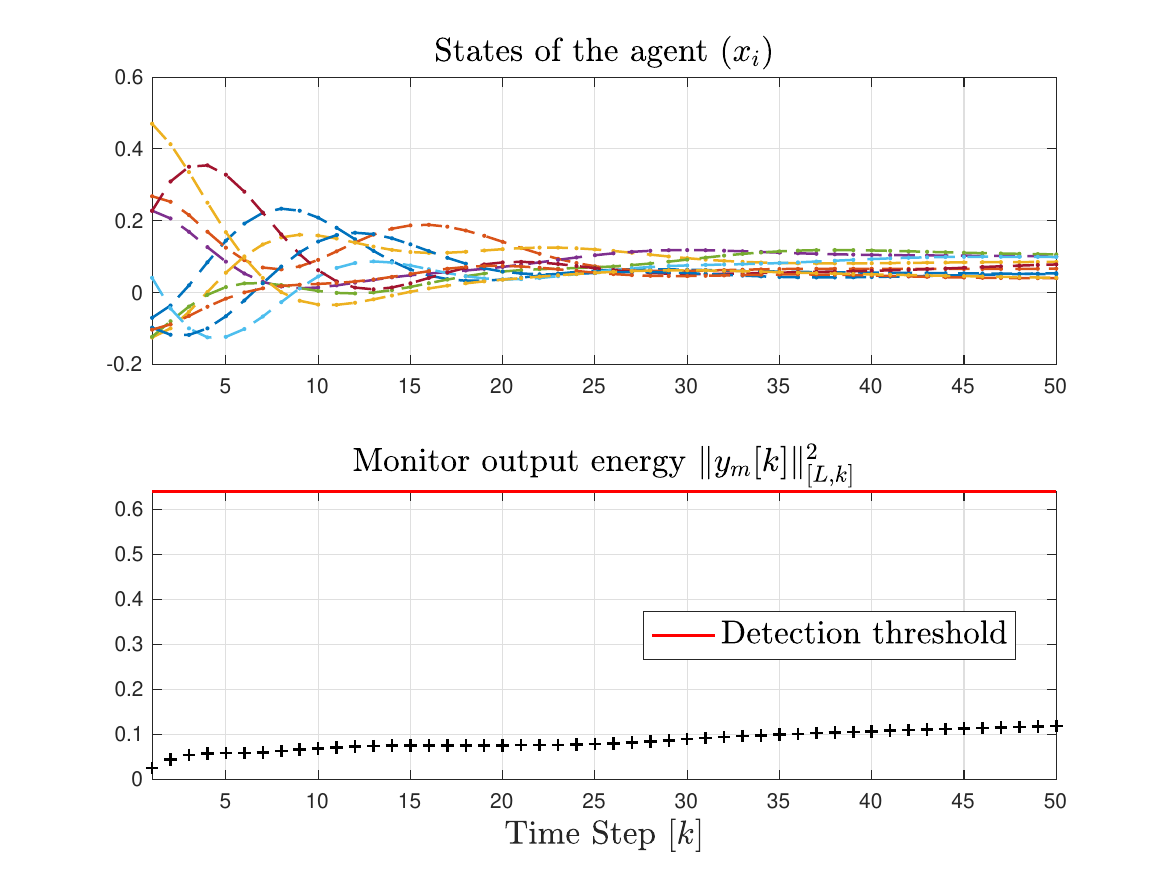}
        \caption{Nominal performance{\color{white}Nominal performanceNominal performance}}
        \label{fig:attack_a}
    \end{subfigure}
    \hfill
    \begin{subfigure}{0.32\textwidth}
        \centering
        \includegraphics[width=1.1\linewidth]{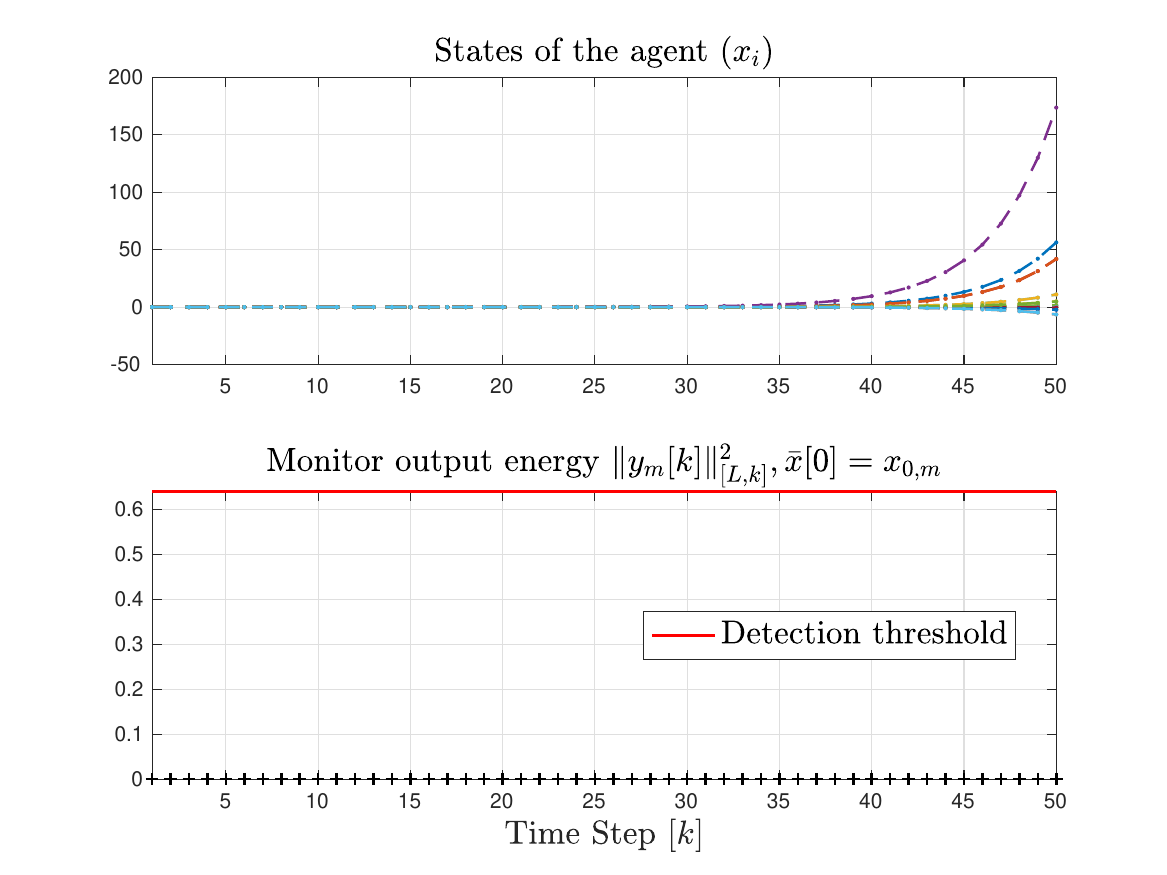}
        \caption{Performance under ZDA with exact initial condition}
        \label{fig:attack_b}
    \end{subfigure}
        \hfill
    \begin{subfigure}{0.32\textwidth}
        \centering
        \includegraphics[width=1.1\linewidth]{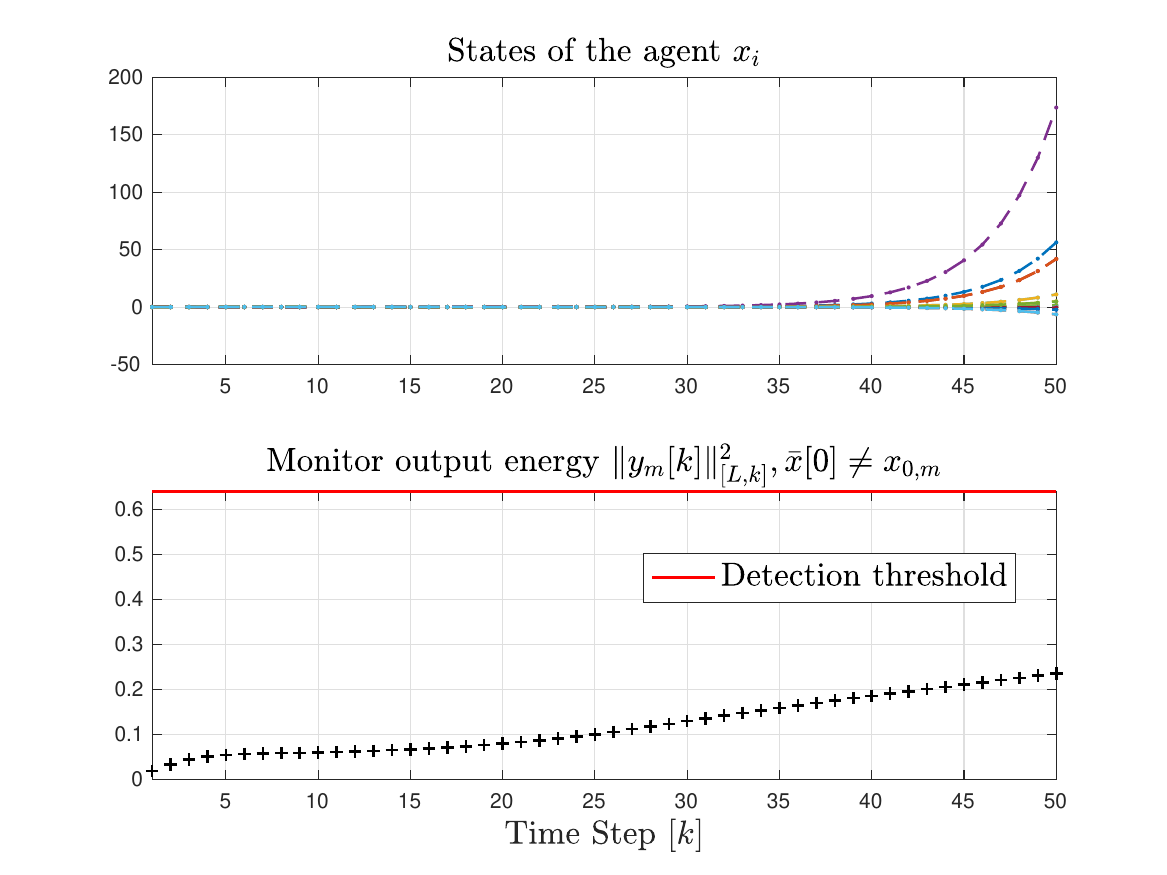}
        \caption{Performance under ZDA with perturbed initial condition}
        \label{fig:attack_c}
    \end{subfigure}
    \caption{Performance of the Wang--Elia algorithm under nominal conditions and under zero dynamics attack (ZDA).}
    \label{fig:attack}
\end{figure*}

\begin{exmp}\label{exmp:1}
Consider a ring graph with $N=10$ nodes. We set the parameters as \(n=1\), \(\alpha = 10^{-6}\), $w=0$ (from \eqref{eq:stealthy}), 
\(Q_i\sim\mathcal{U}[0,\;10], c_{i} \sim \mathcal{U}[0,\;30], \forall\;i \in \{1,2,\dots,N\}\), where \(\mathcal{U}\) denotes the uniform distribution. 
We assume that agent $v_4$ is monitored, while agent $v_8$ is misbehaving. 
Since $w=0$, the monitoring agent observes only the variable $x_4$.

\emph{Nominal Performance:} We select a random initial condition such that $\|\bar{x}[0]\|_2 = 1$. The state trajectories of the agents is shown in Fig.~\ref{fig:attack_a} (top). 
The monitor sets $\kappa = W\check{C}_m \bar{x}[k_1]$, where $k_1 = 100$ is the monitor's estimate of the time for the network to reach approximate consensus.
If this value is unknown, $\kappa_1$ can be chosen to be sufficiently large. 
The norm of the monitoring output $y_m$ is shown in Fig.~\ref{fig:attack_a} (bottom). Then, the detector threshold ($\epsilon$) is determined as follows. Initially, randomized trails are conducted by varying the value of $\bar{x}[0], \|\bar{x}[0]\|_2 = 1$ and the corresponding detector output norm for a specified horizon length is computed. Subsequently, $\epsilon$ is selected such that an alarm is not raised by the detector.

\emph{Attack performance:} Next, consider an adversary that injects an attack of the form~\eqref{eq:attack}. 
The resulting state trajectories and the norm of the monitoring output are shown in Fig.~\ref{fig:attack_b}. 
Due to the nature of the attack, although the states diverge, the monitor output remains unaffected. 
For comparison, Fig.~\ref{fig:attack_c} also illustrates the state trajectories and monitoring output when the initial condition required for the zero dynamics attack is not satisfied, i.e., $\bar{x}[0] \neq x_{0,m}$. 
In this case, the detector output is perturbed, but still no alarm is triggered.
$\hfill \square$
\end{exmp}

%% file: 4_modify_OOG.tex
A common approach to address the ill-posedness of the metric \eqref{eq:OOG} is to bound the energy of the attack input signal \citep{djouadi2014finite,anand2023risk}, ensuring that the resulting performance loss remains finite. However, this approach has two main limitations.

First, the attack energy is generally unknown. Second, even when a bound is available, it provides little guidance for mitigation. For instance, an operator designing the detector threshold $\epsilon$ to minimize the impact in \eqref{eq:OOG} may find the security metric insensitive to $\epsilon$ if any condition in Theorem~\ref{thm:1} holds. This occurs because undetectable attacks do not affect the measurement output. Thus in this section, we introduce methods to resolve the ill-posedness of \eqref{eq:OOG} without relying on assumptions about attack energy.
\subsection{Addressing the zeros at infinity}
As mentioned in Theorem~\ref{thm:1}, the security metric in \eqref{eq:OOG} is unbounded if the relative degree of $\Sigma_m$ is higher than the relative degree of $\Sigma_p$. To address this issue, we consider the security metric \eqref{eq:OOG} under the dynamics \eqref{eq:CL} but with the following modified performance output
\begin{equation}\label{eq:yp}
\tilde{y}_p[k] =  q^{\delta_m-\delta_p} \bar{y}_p[k] = \bar{C}_p \left( \bar{x}[k] q^{\delta_m-\delta_p}\right)
\end{equation}
where $q$ is the backward time-shift operator, and $\delta_m (\delta_p)$ is the relative degree of $\Sigma_m (\Sigma_p)$. In other words, we alter the performance output such that the attack vector has the same relative degree to the performance and detection output. Thus, attacks such as \eqref{eq:attack:rd} cannot cause unbounded performance loss. 

Using the altered performance loss \eqref{eq:yp}, similar to \eqref{eq:CL}, the closed-loop system from the attack input to the modified performance and detection output can be represented as
\begin{equation}\label{eq:CL:mod}
    \begin{aligned}
        \tilde{x}[k+1] &= \tilde{A} \tilde{x}[k] + \tilde{B}a[k] + \tilde{c},\\
        \tilde{y}_p[k] &= \tilde{C}_p \tilde{x}[k],\quad
        \tilde{y}_m[k] = \tilde{C}_m \tilde{x}[k] - \kappa
    \end{aligned}
\end{equation}
where $\tilde{x}$ contains the original state vector $\bar{x}$ from \eqref{eq:CL} along with the time-shifted state trajectories to compute $\tilde{y}_p$ in \eqref{eq:yp}. The measurement signal is represented by $\tilde{y}_m$ to be consistent with the notation. However, for any given attack input the measurement signal from \eqref{eq:CL} and \eqref{eq:CL:mod} are the same. Then we can state the following result.
\begin{prop}\label{prop1}
The relative degree of $\tilde{\Sigma}_m = (\tilde{A},\tilde{B},\tilde{C}_m,0)$ is equal to the relative degree of $\tilde{\Sigma}_p = (\tilde{A},\tilde{B},\tilde{C}_p,0)$. $\hfill \square$
\end{prop}
By introducing artificial delays in \eqref{eq:yp}, we ensure that the relative degree of $\tilde{\Sigma}_p$ matches that of $\tilde{\Sigma}_m$. Consequently, the proof of Proposition~\ref{prop1} follows directly by construction. 

To summarize, in order to address the limitation posed by Theorem~\ref{thm:1}, (1), we consider the security metric \eqref{eq:OOG} under the modified closed-loop dynamics \eqref{eq:CL:mod}. We next propose a tractable approach to compute the value of the security metric under the modified closed-loop dynamics in \eqref{eq:CL:mod}.
\subsection{Convex SDP for the security metric}
Let us assume that enough time has passed for the network to reach approximate consensus, after which the adversary injects an attack. Thus, for the adversary, the network is at consensus when the attack starts. Then, due to linearity, the output of \eqref{eq:CL:mod} can be written as 
\begin{align}
\tilde{y}_{p} &= \tilde{y}_{p,n} + \tilde{y}_{p,a}, \quad
\tilde{y}_{m} = \tilde{y}_{m,n} + \tilde{y}_{m,a},\;\text{where} \\
&\left\{
\begin{aligned}
\tilde{x}_n[k+1] &= \tilde{A} \tilde{x}_n[k] + \tilde{c}, \; \tilde{x}_n[0] = \tilde{x}_0 \\
\tilde{y}_{p,n}[k] &= \tilde{C}_p \tilde{x}_n[k], \;
\tilde{y}_{m,n}[k] = \tilde{C}_m \tilde{x}_n[k] - \kappa
\end{aligned}
\right\} \\
&\left\{
\begin{aligned}\label{eq:eq2}
\tilde{x}_a[k+1] &= \tilde{A} \tilde{x}_a[k] + \tilde{B} a[k],\; \tilde{x}_a[0] = \mathbf{0} \\
\tilde{y}_{p,a}[k] &= \tilde{C}_p \tilde{x}_a[k],\;\tilde{y}_{m,a}[k] = \tilde{C}_m \tilde{x}_a[k]
\end{aligned}
\right\}
\end{align}
where the subscripts $n$ and $a$ denotes the nominal output and the attacked outputs, respectively, and $\tilde{x}_0$ denotes the consensus value. Then, under the performance output \eqref{eq:CL:mod}, the security metric \eqref{eq:OOG} can be similarly written as
\begin{equation}\label{eq:MOOG}
\sup_{a \in \ell_{2e}} \left\{ \Vert \tilde{y}_{p,a} \Vert_{\ell_2, [1, L]}^2 \vert \; \Vert \tilde{y}_{m,a}\Vert_{\ell_2, [1, L]}^2 \leq \epsilon,\; \eqref{eq:eq2} \right\}
\end{equation}
whose value is denoted by  $\tilde{\gamma}(v_m,v_a,K)$, where $v_m$ is the agent monitored, and $v_a$ is the agent under attack. Note that the matrix $\tilde{B}$, $\tilde{C}_m$, and $\tilde{A}$ depends on $v_a$, $v_m$, and $K$, respectively. Next we provide a tractable SDP to determine the value of \eqref{eq:MOOG}.
\begin{thm}
The value of the security metric \eqref{eq:MOOG} is given by the value of the following SDP
\begin{equation}\label{eq:SDP}
\tilde{\gamma}(v_m,v_a,K) = \left\{ \begin{aligned}
        \min_{\gamma, P} & \quad  \epsilon \gamma \\
        \text{s.t.} & \quad \begin{bmatrix}
            R & \tilde{A}^TP\tilde{B}\\
            \tilde{B}^TP\tilde{A} & \tilde{B}^TP\tilde{B}
        \end{bmatrix} \preceq 0\\
        & \quad \gamma \geq 0, P=P^T, P \succeq 0
    \end{aligned}\right.
\end{equation}
where $R\triangleq \tilde{A}^TP\tilde{A}-P + \tilde{C}_p^T\tilde{C}_p - \gamma \tilde{C}_m^T\tilde{C}_m$
\end{thm}
\begin{pf}
For any given $L \in \mathbb{Z}^{+}$, the optimization problem \eqref{eq:OOG} is equivalent to its dual problem
\begin{equation}\label{eq7}
\inf_{\gamma}\left\{ \epsilon\gamma \Big| ||\bar{y}_{p}||_{\ell_2,[1,L]}^2 - \gamma ||\bar{y}_{m}||_{\ell_2,[1,L]}^2 \leq 0 \;\; \forall a \in \ell_{2e} \right\}.
\end{equation}
The duality gap was shown to be zero in \citep{anand2023risk}. Using dissipative system theory \cite[Chapter 5.7]{moylan2014dissipative}, \eqref{eq7} can be re-written as \eqref{eq:SDP} which concludes the proof. $\hfill \blacksquare$
\end{pf}
Efficient numerical solvers exist to compute the solution of SDPs (see \cite{huang2022solving} and references therein). Additionally, the value of the SDP is independent of the horizon length $L$. Thus if the operator is interested in quantifying the performance loss for a given horizon length, the value of the SDP acts as an upper bound.
\subsection{Addressing the invariant zeros} As mentioned in Theorem~\ref{thm:1}, if $\Sigma_m$ has an unstable zero which is not a zero of $\Sigma_p$, then the security metric \eqref{eq:OOG} or \eqref{eq:MOOG} is unbounded. This is because, the unstable zeros are insensitive to the delays introduced in \eqref{eq:CL:mod}. Thus we aim to introduce a modification to the metric \eqref{eq:MOOG} such that its value is bounded in the presence of unstable zeros. 

Firstly, by numerical experiments, we found that in general if $w \in (0,1)$, then $\Sigma_m$ has no unstable zeros. In other words, if the variables $x$ and $v$ are both monitored, then there are no unstable zeros (similar to \cite{shaaban2024cyber}). We explain this in Example~\ref{exmp:2}.
\begin{exmp}\label{exmp:2}
Consider a ring graph with $N=10$ agents where $v_8$ is malicious, and $v_4$ is the monitoring agent. When $w=0$, there exists an unstable zero for $\Sigma_m$ at $\lambda = 1.4$. Similarly when $w=1$, the unstable zero of $\Sigma_m$ is at $\lambda = 2.6$. However, if $w=0.5$, there are no zeros for $\Sigma_m$. This observation remain consistent across different graph structures and step sizes.$\hfill \square$
\end{exmp}
Although we have found via extensive numerical simulations that the network has no unstable invariant zeros when $w \in (0,1)$, we cannot guarantee the non-existence of an unstable zero theoretically (see Remark~\ref{rem:attack:powerful}). Thus, we next provide a theoretical modification to the security metric such that the metric is well posed in the presence of an unstable zero.
\begin{thm}\label{thm:zero:unit}
Let the set of zeros of $\bar{\Sigma}_m$ and $\bar{\Sigma}_p$ be denoted by $\mathcal{Z}_m \triangleq \{\lambda_{m,1},\dots,\lambda_{m,{\zeta_m}}\}$ and $\mathcal{Z}_p \triangleq \{\lambda_{p,1},\dots,\lambda_{p,{\zeta_p}}\}$, respectively. Suppose that $\vert \lambda_{m,1}\vert >1$, $\vert \lambda_{p,1}\vert >1$, $\vert \lambda_{m,1}\vert \neq \vert \lambda_{p,1}\vert$, and $\vert \lambda_{m,j}\vert <1, \vert \lambda_{p,j}\vert <1$ for all $j \in \{2, \dots,{\zeta_m}\}$ and $q \in \{2, \dots,{\zeta_p}\}$. Then the value of \eqref{eq:SDP} is bounded if the constraint $P \succeq 0$ is removed.
\end{thm}
\begin{pf}
When the constraint $P \succeq 0 $ is removed, using cyclo dissipativity \cite[Proposition~2]{teixeira2019optimal}, we can conclude that the value of \eqref{eq:SDP} is bounded if there are no zeros on the unit circle (similar to the proof of \cite[Lemma 4.5]{anand2023risk_tac}). Since by construction, there are no zeros on the unit circle, the proof concludes. $\hfill \blacksquare$
\end{pf}

The SDP \eqref{eq:SDP} with the constraint $P \succeq 0$ considers all trajectories that starts from $\tilde{x}_{a}[0]=0$. If the constraint is dropped, and is replaced with a only symmetric matrix $P=P^T$, then the SDP considers all trajectories that start from $\tilde{x}_a[0]=0$ and ends at $\tilde{x}_a[L]=0$; thanks to cyclo-dissipativity \cite[Chapter 5.7]{moylan2014dissipative}. In other words, if the constraint $P \succeq 0$ is ignored, we consider an adversary that satisfies the constraint $\tilde{x}[L]=0$. If the adversary does not want to trigger an alarm after the attack horizon, then considering a cyclo-dissipative attack is rational and scalable {(here scalability results from dropping the constraint $P \succ 0$)}.
\begin{exmp}\label{exmp:3}
Consider a ring graph with $N=30$ nodes. We set the parameters as \(n=1\), \(\alpha = 10^{-1}\), $w=0.5$ (from \eqref{eq:stealthy}), 
\(Q_i\sim\mathcal{U}[0,\;1], c_{i} \sim \mathcal{U}[0,\;2], \forall\;i \in \{1,2,\dots,N\}\)
and where \(\mathcal{U}\) denotes the uniform distribution. We assume that agent $v_4$ is monitored, while agent $v_3$ is malicious. The relative degree of $\Sigma_m$ is $1$ and the relative degree of $\Sigma_p$ is $0$. Thus the condition in Theorem \ref{thm:1} is satisfied, and the impact is unbounded. We alter the performance output as in \eqref{eq:yp} and compute the security metric by solving \eqref{eq:SDP} and obtain $\tilde{\gamma}^* =6287.84$.

Here the SDPs are solved using Matlab 2024a with YALMIP \citep{lofberg2004yalmip} and MOSEK \citep{aps2019mosek}. On average, it takes approximately $17 \;\mbox{s}$ to solve the SDP without the constraint $P \succeq 0$. However, with the constraint the solver takes $28 \;\mbox{s}$. Thus, removing the constraint can affect the salability of the SDP for large $N$. $\hfill \square$
\end{exmp}
\begin{rem}\label{rem:attack:powerful}
\SCB{Example~\ref{exmp:2} indicated that the attack cannot cause unbounded consensus deviation when both states are monitored, i.e., $\tilde{y}_m \in \mathbb{R}^{2n}$. In this case, the attack dimension is smaller than that of the detector. However, if distinct attacks are injected into $x$ and $z$ (see Remark~\ref{rem:attack:different}), the attack dimension matches the detector dimension, and the observation in Example~\ref{exmp:2} may no longer hold. In such cases, the result of Theorem~\ref{thm:zero:unit} proves useful.} $\hfill \triangleleft$
\end{rem}

%% file: 5_Design.tex
In the previous section, we presented a method to quantify the deviation in consensus caused by a malicious agent. Such metrics, which quantify the consensus deviation, can also be used for design purposes. In this section, we present two design approaches: (a) optimal monitor design \citep{nguyen2024security}, and (b) optimal edge design. 
\subsection{Optimal monitor design}
Let us define the set $\mathcal{A}$, which contains the set of nodes that can probably be malicious. In other words, the operator has a probability belief over the set of agents that can be under attack. {Let us denote the probability of a node $v_a \in \mathcal{A}$ being under attack is $\varphi_{v_a}$. It holds that $\sum_{v_a \in \mathcal{A}} \varphi_{v_a} =1$.}
Given that some of the nodes are under attack, the operator aims to optimally select the monitor node $v_m$. In other words, the operator is interested in solving the following optimization problem
\begin{equation}\label{eq:design:1}
\arg \;\inf_{v_m}\; \sum_{v_a \in \mathcal{A}} \varphi_{v_a} \tilde{\gamma}(v_m,v_a,K)
\end{equation}

\begin{figure}
    \centering
    \includegraphics[width=0.95\linewidth]{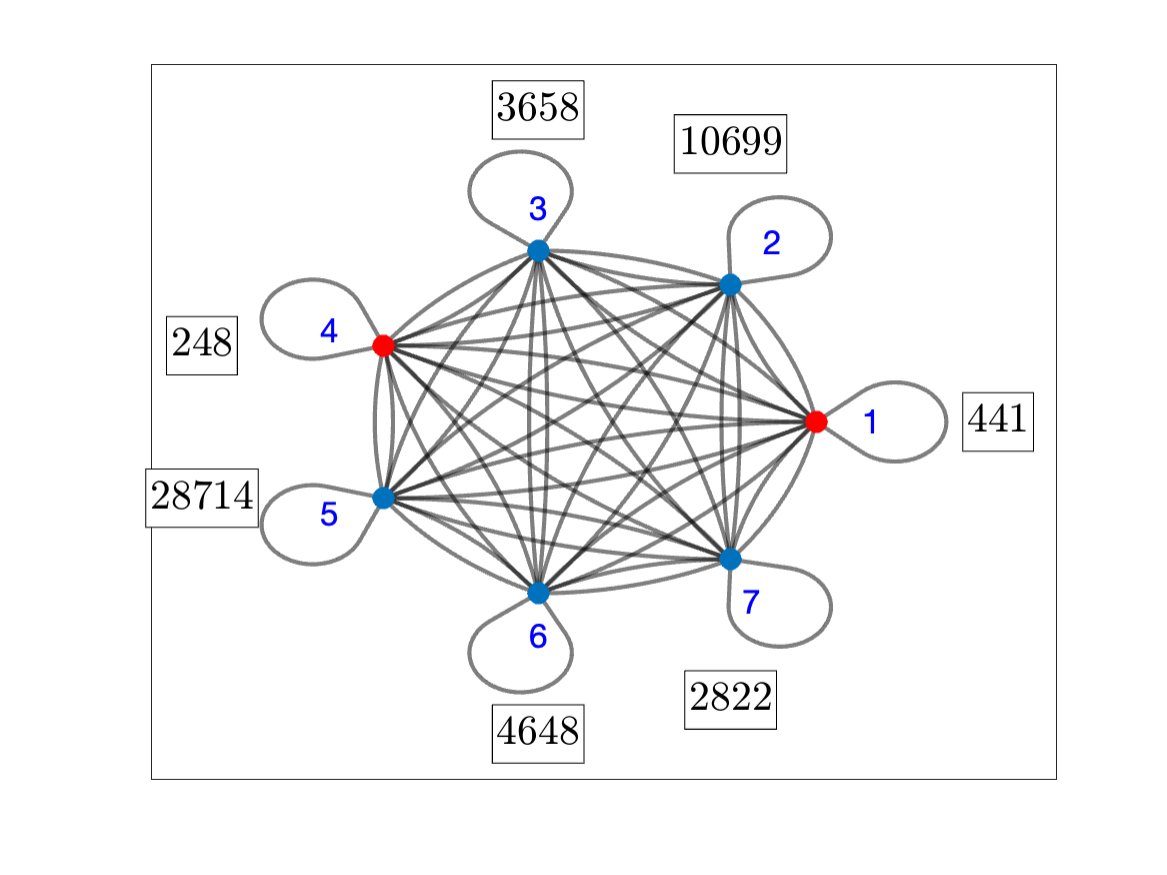}
    \caption{Graph structure considered in Example~\ref{exmp:design1}. Edges are represented by solid black lines, and the red nodes indicate nodes possibly under attack. The value of \(\sum_{v_a \in \mathcal{A}} \tilde{\gamma}(v_m, v_a,K)\) corresponding to each candidate monitoring node \(v_m\) is shown next to the node.}
    \label{fig:design1}
\end{figure}

\renewcommand{\algorithmicrequire}{\textbf{Input:}}
\renewcommand{\algorithmicensure}{\textbf{Output:}}
\newcommand{\INITIALIZE}{\STATE\textbf{Initialize:} }
\begin{algorithm}[t]
\caption{Optimal Monitor Design}
\label{alg:exhaust}
\begin{algorithmic}[1]
\REQUIRE $\mathcal{V}$, $\mathcal{A}$, $\varphi_{v_a}\; \forall v_a \in \mathcal{A}$, and empty list $\mathcal{L}$
\ENSURE Optimal cost $\gamma^*$ and monitor $v_m^*$
\INITIALIZE $\gamma^* \gets \infty$, $v_m^* \gets \emptyset$
\WHILE{$|\mathcal{L}| < N$}
    \STATE Select a node $v_m \in \mathcal{V}$ and delete it from $\mathcal{V}$
    \STATE $\mathcal{L} \gets \mathcal{L} \cup v_m$, $\beta \gets 0$
    \FOR{$i = 1$ to $|\mathcal{A}|$}
        \STATE Select $i^{\text{th}}$ element of $\mathcal{A}$ and denote by $v_a$
        \STATE Compute $\tilde{\gamma}(v_m,v_a,K)$
        \STATE $\beta \gets \beta + \varphi_{v_a} \tilde{\gamma}(v_m,v_a,K)$
    \ENDFOR
    \STATE If $\beta < \gamma$, then $\gamma^* \gets \beta$, $v_m^* \gets v_m$
\ENDWHILE
\RETURN $(\gamma^*, v_m^*)$
\end{algorithmic}
\end{algorithm}
In words, the term $\sum_{v_a \in \mathcal{A}} \varphi_{v_a} \tilde{\gamma}(v_m,v_a,K)$ captures the expected, over the number of attack nodes, maximum deviation in consensus caused when agent $v_m$ is monitored. Thus, we aim to seek the agent $v_m$ that causes the least deviation in consensus when one of the agents in $\mathcal{A}$ is under attack. We describe an algorithm that optimally solves the design problem \eqref{eq:design:1} using an SDP in Algorithm~\ref{alg:exhaust}. 
\begin{exmp}\label{exmp:design1}
Consider the graph depicted in Fig.~\ref{fig:design1}. The parameters are set as \(n = 1\), \(\alpha = 10^{-6}\), \(w = 0.5\) (from \eqref{eq:stealthy}), 
with \(Q_i \sim \mathcal{U}[0,10]\) and \(c_i \sim \mathcal{U}[0,2]\) for all \(i \in \{1,2,\dots,N\}\). Let \(\mathcal{A} \triangleq \{v_1, v_4\}\). Here, we assume that all nodes are equally likely to be under attack; hence, the term $\varphi_{v_a}$ is omitted. The value of \(\sum_{v_a \in \mathcal{A}} \tilde{\gamma}(v_m, v_a)\) for each candidate monitoring node \(v_m\) is shown next to the corresponding node in Fig.~\ref{fig:design1}, from which it is also evident that monitoring node \(v_4\) is optimal. $\hfill \blacksquare$
\end{exmp}
\begin{rem}
{Most of the existing literature assumes that a node under attack cannot be monitored. However, if the operator suspects that agent $v_a$ may be compromised, it is reasonable to monitor its output directly. This decision is sensible, as it represents a \emph{preemptive} action taken before any actual attack has possibly occurred.} $\hfill \triangleleft$
\end{rem}
\subsection{Optimal edge design}
Selecting the monitor node is one design choice. If the monitor node cannot be altered, another way to alter the design performance is to change the graph structure. In particular, let us consider an operator who is interested in adding \emph{one} edge to the network such that the value of the security metric is lower. Let $\mathcal{F}=\{e_{f,1},e_{f,2},\dots\}$ represent the set of all edges that the operator can add and let $\phi_i$ represent the cost of adding an edge $e_{f,i} \in \mathcal{F}$ to the network. Then the operator is interested in adding an edge by solving the following design problem 
\begin{equation}\label{eq:design:2}
\arg \;\inf_{e_{f,i} \in \mathcal{F}}\; \phi_i + \sum_{v_a \in \mathcal{A}} \varphi_{v_a}\tilde{\gamma}\left(v_m,v_a,K(e_{f,i})\right) 
\end{equation}
In words, the total cost that the operator aims to minimize is the sum of the edge cost and the term
$\sum_{v_a \in \mathcal{A}} \varphi_{v_a} \tilde{\gamma}\left(v_m,v_a,K(e_{f,i})\right)$ which represents the average consensus deviation caused when $v_m$ is monitored, and edge $e_{f,i}$ is added to the network. We describe an algorithm that optimally solves the design problem \eqref{eq:design:2} using an SDP in Algorithm~\ref{alg:exhaust2}. 
\begin{rem}
{Here, we consider an operator that adds an edge; however, Algorithm~\ref{alg:exhaust2} can also be used to optimally remove an edge. In this case, $\phi_i$ represents the cost associated with removing the edge $e_{f,i}$, and $K(e_{f,i})$ denotes the adjacency matrix obtained after removing that edge.} $\hfill \triangleleft$
\end{rem}

\begin{figure}
    \centering
    \includegraphics[width=0.95\linewidth]{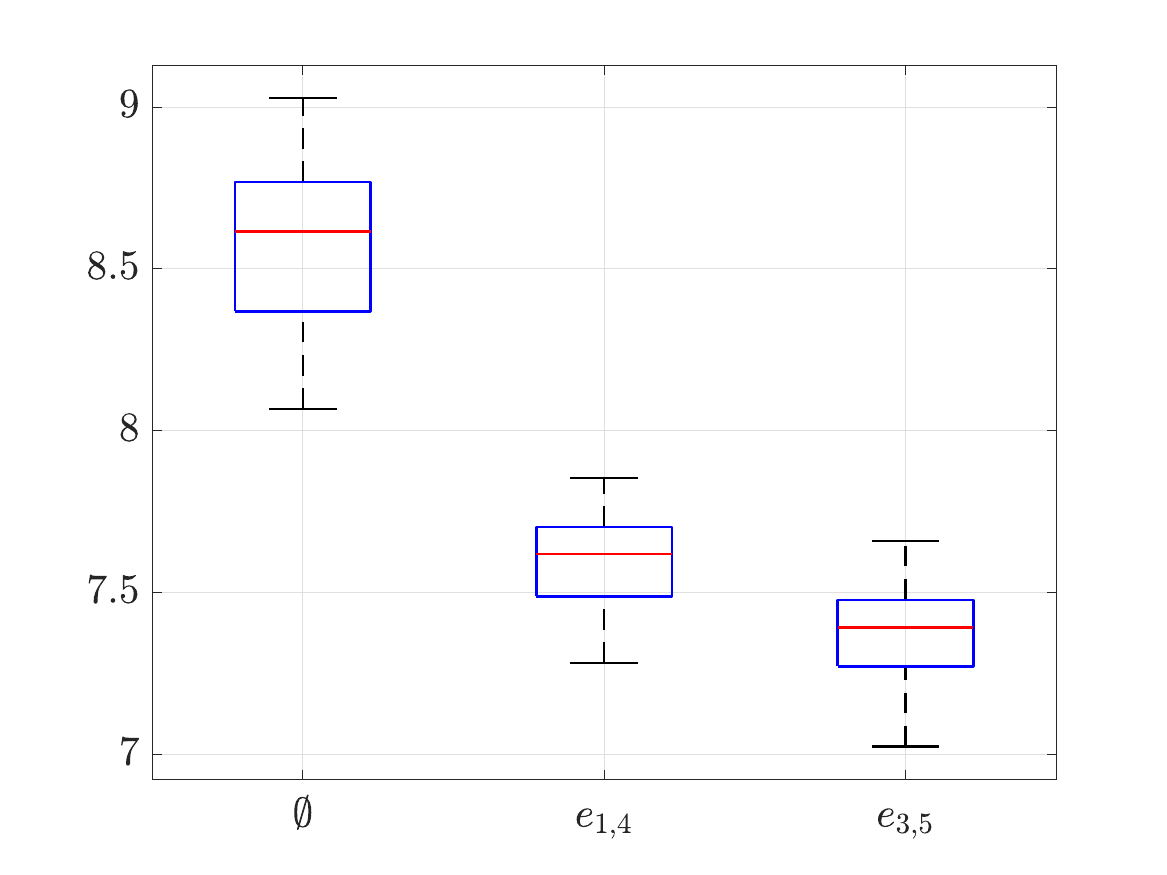}
    \caption{Box plot of $\log\left(\phi_i + \sum_{v_a \in \mathcal{A}} \tilde{\gamma}\left(v_3,v_1,K(e_{f,i})\right)\right)$ for a ring graph when the edges denoted in the x-axis are added. On each box, the central mark indicates the median, and the bottom and top edges of the box indicate the $25^{th}$ and $75^{th}$ percentiles, respectively. The whiskers extend to the most extreme data points.}
    \label{fig:design2}
\end{figure}

\begin{algorithm}[t]
\caption{Optimal Edge Design}
\label{alg:exhaust2}
\begin{algorithmic}[1]
\REQUIRE $\mathcal{E}$, $\mathcal{F}$, and empty list $\mathcal{L}$
\ENSURE Optimal cost $\gamma^*$ and edge $e^*$
\INITIALIZE $\gamma^* \gets \infty$, $e^* \gets \emptyset$
    \STATE Compute $\sum_{v_a} \varphi_{v_a}\tilde{\gamma}(v_m,v_a,K(\emptyset))$
    \STATE $\gamma^* \gets \sum_{v_a} \varphi_{v_a}\tilde{\gamma}(v_m,v_a,K(\emptyset))$, $e^* \gets \emptyset$
\WHILE{$|\mathcal{L}| \leq |\mathcal{F}|$}
    \STATE Select an edge $e_{f,i} \in \mathcal{F}$ and delete it from $\mathcal{F}$
    \STATE $\mathcal{L} \gets \mathcal{L} \cup e_{f,i}$, $\beta \gets \phi_{i}$
    \STATE Compute the adjacency matrix $K$ with the added edge $e_{f,i}$
    \FOR{$i = 1$ to $|\mathcal{A}|$}
        \STATE Select $i^{\text{th}}$ element of $\mathcal{A}$ and denote by $v_a$
        \STATE Compute $\tilde{\gamma}(v_m,v_a,K(e_f))$
        \STATE $\beta \gets \beta + \varphi_{v_a}\tilde{\gamma}(v_m,v_a,K(e_f))$
    \ENDFOR
    \STATE If $\beta < \gamma$, then $\gamma^* \gets \beta$, $e^* \gets e_{f,i}$
\ENDWHILE
\RETURN $(\gamma^*, e_{f,i}^*)$
\end{algorithmic}
\end{algorithm}
\begin{exmp}\label{exmp:design2}
Consider a ring graph with $N=5$ nodes where the parameters are set as \(n = 1\), \(\alpha = 10^{-1}\), \(w = 0.5\) (from \eqref{eq:stealthy}), 
with \(Q_i \sim \mathcal{U}[0,5]\) and \(c_i \sim \mathcal{U}[0,3]\) for all \(i \in \{1,2,\dots,N\}\). Let \(v_a = v_1\), and $v_m=v_3$, and $\mathcal{F}=\{e_{1,4}, e_{3,5}\}$, and $\phi_{e_{ij}}=i \times j \times 10$. For each edge $e_{f,i}$, we compute the value of $\phi_i +\tilde{\gamma}\left(v_3,v_1,K(e_{f,i})\right)$ for $\bar{N}=100$ randomized experiments. The values are depicted in Fig.~\ref{fig:design2}. We can infer from the figure that it is optimal to add the edge $e_{3,5}$ since it reduces the security metric the most. $\hfill \blacksquare$
\end{exmp}

%% file: 6_Discussion.tex
In summary, this paper provided a method to quantify the security of the gradient tracking algorithms for quadratic optimization problems when one of the agents is under attack. In Section~\ref{sec:design}, we also used the quantified security towards designing robust network architectures. In this section, we extend the security analysis when each agent has a nonlinear, polynomial (in the state) objective function. Then the objective of DO is to find the value of $x$ that minimizes the following cost 
\begin{equation}
\sum_{i=1}^{N} f_i(x),  f_i(x) \triangleq A_iZ(x)
\end{equation}
where $Z(x)$ is a polynomial basis function in $x$. As mentioned before, since each agent has only access to $f_i(x)$, the agents update their estimate of $x$ using the Wang -- Elia algorithm, which can be written compactly as follows 
\begin{equation}\label{eq:CL:poly}
\begin{aligned}
\bar{x}[k+1] &= \bar{A}Z(\bar{x}) + \bar{B}a[k]\\
\bar{y}_p[k] &= C_p Z(\bar{x}),\;\; \bar{y}_m[k] = C_m Z(\bar{x})
\end{aligned}
\end{equation}
where $\bar{x}$ contain the stacked states of the agents $\{x_i,z_i\}$, and $\bar{y}_p$ and $\bar{y}_m$ are the performance and measurement outputs, respectively. We next quantify the maximum deviation in consensus caused by a stealthy adversary, which is given by the value of the following optimization problem 
\begin{equation}\label{OOG:poly:pri}
\gamma_{PL} \triangleq \sup_{a \in \ell_{2e}} \left\{\; \Vert \bar{y}_p \Vert_{\ell_2, [1, L]}^2 \vert \left\Vert \bar{y}_m\right\Vert_{\ell_2, [1,L]}^2 \leq \epsilon,\;\eqref{eq:CL:poly} \right\}
\end{equation}

where $\gamma_{PL}$ represents the deviation caused on a PolynomiaL objective function. We next provide a Sum-Of-Squares (SOS) formulation to determine the value of \eqref{OOG:poly:pri}. Also, efficient numerical solvers exist to compute the solution of the SOS program \citep{prajna2002introducing}. After presenting the results, we demonstrate the efficacy of the SOS program via a numerical example. 
\begin{prop}
The value of \eqref{OOG:poly:pri} is bounded above by the value of the following SOS program
\begin{equation}\label{eq:SOS}
\begin{aligned}
\min_{\gamma, D} & \;\; \epsilon \gamma\\
\text{s.t.} &\;\; g(\bar{x},\gamma) \;\text{is SOS}, \gamma \geq 0, D\succ 0\\
&\;\; g(\bar{x},\gamma) \triangleq -Z(\bar{x}[k+1])^TDZ(\bar{x}[k+1])+\\
& \qquad \quad Z^T(\bar{x}[k])(\gamma C_m^TC_m-C_p^TC_p+D)Z(\bar{x}[k])
\end{aligned}
\end{equation}
\end{prop}
\begin{pf}
Using duality, we can recast the optimization problem \eqref{OOG:poly:pri} as
\begin{align}\label{s1}
\inf \{\epsilon \gamma |\left\Vert \bar{y}_p\right\Vert_2^2 - \gamma \left\Vert \bar{y}_p\right\Vert \leq 0.\}
\end{align}
Consider a storage function of the form $S(\bar{x})= Z(\bar{x})DZ(\bar{x})$, and the supply rate of the form  $w(\cdot) \triangleq Z^T(\bar{x}[k])(\gamma C_m^TC_m-C_p^TC_p)Z(\bar{x}[k])$. Then, it follows from \citet[Proposition 11]{papachristodoulou2005tutorial} that \eqref{s1} is equivalent to \eqref{eq:SOS}. However, it is not necessary for the storage function to be polynomial. In other words, there can exist a different storage function $\tilde{S}(\bar{x}[k])$ that makes \eqref{eq:CL:poly} dissipative w.r.t $w(\cdot)$ with a lower value of $\gamma$. Thus, the value of $\gamma_{PL}$ obtained by solving the SOS program is an upper bound on the value of \eqref{OOG:poly:pri}. This concludes the proof. $\blacksquare$
\end{pf}
\begin{exmp}
Consider a ring graph with $N=2$, \(n=1\), \(\alpha = 10^{-4}\), and \(f_i(x) = \frac{1}{2}x^2-\frac{1}{6}x^3+\frac{1}{8}x^4\). We assume that agent $v_2$ is monitored, while agent $v_1$ is misbehaving. We solve the SOS program \eqref{eq:SOS} using Matlab 2024a with YALMIP \citep{lofberg2004yalmip} and MOSEK \citep{aps2019mosek}. We determine the value of $\gamma_{LP} = 4.02 \times 10^{-6}$. In general, this means that the adversary can only induce an error of the order of $10^{-6}$; however, the significance of the value depends on the accuracy expected by the operator. 
\end{exmp}